\documentclass[aps,prd,nofootinbib]{revtex4}

\usepackage{graphicx}
\usepackage{amsmath}   
\usepackage{amssymb}

\usepackage{epsfig}
\usepackage{color}

\usepackage{amsmath}

\usepackage{amssymb}

\begin{document}
\title{Energy conditions and gravitational baryogenesis in $f(R, {\cal R})$ gravity}
\author{K. Atazadeh}\email{atazadeh@azaruniv.ac.ir}
\author{S. Golsanamlou}\email{samira.golsanamlou@azaruniv.ac.ir}
\affiliation{Department of Physics, Azarbaijan Shahid Madani University, Tabriz, 53714-161 Iran}

\begin{abstract}
In this work, first we examine the energy conditions in the context of the generalized metric-Palatini hybrid gravity, known as $f(R, {\cal R})$ gravity. We show that for the proposed model in this study, {\it i.e.}  $ f(R, {\cal R})= R +\alpha{\cal R}^{n}  $, one of the four fundamental energy conditions, specifically the strong energy condition, does not hold for some values of $n$. Therefore, it seems that hybrid gravity can provide a model for the accelerated expansion of the universe.
In continuation of completing our study in this work, we try to analyze the impact of hybrid metric-Palatini gravity on the gravitational baryogenesis process. The hybrid metric-Palatini model combines two gravitational theories that allow for a more detailed examination of the behavior of space-time and its interaction with matter. This combination is critical in the early radiation-dominant universe, where unusual gravitational effects may play a key role in generating baryonic asymmetry and the production of baryons and anti-baryons.

\end{abstract}

\maketitle
\date{}
\section{Introduction}

One of the biggest challenges currently faced by standard cosmology is describing the accelerated expansion of the universe. This acceleration has been confirmed by various cosmological observations, such as the study of the redshift curves of Type Ia supernovae, the examination of the large-scale structure of the universe, and observations of cosmic microwave background radiation (CMB) \cite{1,2,3,4}. Various methods have been proposed to elucidate this acceleration. One such method involves the introduction of a mysterious cosmic fluid known as dark energy, characterized by its negative pressure. Additionally, modified gravity theories are regarded as potential explanations for the perplexing characteristics of dark energy \cite{5}.
\\
The generalized gravity models involve extending the Einstein-Hilbert action into an arbitrary function of the Ricci scalar curvature, a popular one of them is $ f(R)$ gravity. In this model, the field equations are replaced with additional terms from the Ricci scalar curvature instead of Einstein's general relativity (GR) equations. Although solving these equations leads to accelerated solutions for the universe without the need to introduce dark energy, the freedom in constructing various forms of the function $f(R)$ creates a challenge in constraining it from both theoretical and observational perspectives.
Thus, one way to impose constraints on the generalized gravity theories is by applying energy conditions.
\\
In this direction, we study the energy conditions in the generalized metric-Palatini theory of gravity characterized by the function of $ f(R, {\cal R})$, in which $R$ is the metric scalar curvature and ${\cal R}$ is the Palatini scalar curvature on the manifold.
Specific conditions must be applied to the energy-momentum tensor $ T_{\mu\nu} $, known as energy conditions, to depict a realistic distribution of matter. There are four fundamental energy conditions in GR that they known as the null energy condition (NEC), the weak energy condition (WEC), the dominant energy condition (DEC), and the strong energy condition (SEC). They play significant roles in studying singularity theorems, black hole thermodynamics, and other areas in physics \cite {18}. Various studies in the context of metric-Palatini hybrid gravity have been investigated, see Refs. \cite{6,7,8,9,10,11,12,13,14,15,16,17}.
\\
In this work, we first study the energy conditions scheme in hybrid metric-Palatini gravity for a homogeneous and isotropic universe, specifically in the Friedmann-Robertson-Walker (FRW) background. In the continuation to complete our study in the last section, we consider the gravitational baryogenesis for the hybrid metric-Palatini $ f(R,{\cal R}) $ gravity.
Baryogenesis is the process in the early universe that led to an asymmetry between matter and antimatter, resulting in more baryons (such as protons and neutrons) than antibaryons \cite{1b}. This baryon-antibaryon asymmetry is a crucial issue in cosmology, as it has profound implications for the formation of large-scale structures like galaxies and stars. Observational evidence from the CMB and the Big Bang Nucleosynthesis (BBN) supports the existence of this imbalance, which eventually led to the material universe we observe today \cite{2b,3b,4b}. Despite extensive research, the underlying cause of the baryonic asymmetry remains an unresolved mystery \cite{5b}.\\
Gravitational baryogenesis is a theoretical framework proposed to explain the baryon asymmetry observed in the universe. This mechanism suggests that gravity, particularly at very small scales or in the early universe, can play a crucial role in the generation of baryon-antibaryon asymmetry.
In accordance with the Sakharov criteria \cite{7b}, gravitational baryogenesis responds to these criteria by suggesting that the gravitational field present in the early universe may initiate processes that contravene baryon number conservation and CP symmetry, especially during epochs like the radiation-dominated era when the universe was undergoing rapid cooling.
The term for the imbalance of matter and antimatter in gravitational baryogenesis is expressed as \cite{6b}
\begin{equation}
\label{0}
\frac{1}{M_{*}^2}\int d^4x\sqrt{-g}(\partial_{\mu}R)J^{\mu}\, .
\end{equation}
This term can be resulted from the higher-order interactions in the gravitational physics. In the equation (\ref{0}), $M_*$ is the cutoff scale of the underlying effective gravitational theory, $J^{\mu}$ is the baryon flux describing the number of baryons and their direction of motion in space-time, $\partial_{\mu}R$ is the fourth derivative of the Ricci scalar representing local variations of spacetime curvature.
In the context of modified gravity models, gravitational baryogenesis provides a mechanism for baryon production that satisfies these conditions \cite{8b,9b,10b,11b,12b,13b}.\\
The purpose of this study is to derive the energy conditions directly from the effective energy-momentum tensor approach through the transformations  $\rho\longrightarrow \rho^{eff} $ and $ p\longrightarrow p^{eff} $. The energy conditions' inequalities were computed in relation to the cosmological parameters. We examined a particular form of $f(R, {\cal R})$ gravity, namely  $ f(R, {\cal R}) = R + \alpha {\cal R}^n $. We showed that for certain values of $n$, the WEC is fulfilled.
We illustrate that the hybrid metric-Palatini gravity model effectively captures the dynamics of the universe across various epochs. In particular, for both the matter-dominated and radiation-dominated periods, the derived values closely match the observed value of approximately $10^{-11}$, thereby validating the model as a robust framework for exploring cosmic phenomena.
The alignment between theoretical predictions and observational data indicates that the hybrid metric-Palatini model may significantly contribute to future cosmological research. Additionally, the modified baryon-to-entropy ratio obtained from this model is consistent with cosmological observations, underscoring its potential to deepen our understanding of baryogenesis and to explore gravitational processes beyond the standard GR framework.
Gravitational baryogenesis offers a compelling explanation for the origin of the baryon asymmetry and its contribution to the formation of the material universe we observe today. Thus, the hybrid metric-Palatini gravity model that combines metric and Palatini approaches (torsionless connection, $\hat{\Gamma}^{\lambda}_{\mu\nu}$) and allows gravity to operate more accurately in the early universe, and simulates unusual effects of gravity that the standard model cannot explain.
\\
The paper is organized as follows: The field equations in the $ f(R, {\cal R})$ gravity and the generalized Friedmann equations are developed in section \textsc{II}. The energy conditions are presented in section \textsc{III} and in this section we take a functional form for $f(R, {\cal R})$ to study the energy conditions for the proposed model. Section \textsc{IV} covers field equations of $f(R, {\cal R})$ gravity in the context of scalar-tensor theories and also we consider of the energy conditions in this model. We consider the impact of hybrid metric-Palatini gravity on the gravitational baryogenesis process in section \textsc{V}. Finally, we offer our conclusions in the last section.


\section{Friedman equations in $f(R, {\cal R})$ gravity}
In modified gravity theories $ f(R, {\cal R}) $, the field equations are derived from the modified action. The four-dimensional action for hybrid metric-Palatini gravity is given as follows \cite{19}

\begin{equation}\label{1}
S=\frac{1}{2\kappa}\int d^4{x}\sqrt{-g}[R+{f({\cal R})}]+S_{m},
\end{equation}
where $\kappa=8\pi G$ and $G$ is gravitational constant. In the action \eqref{1}, $ R $  is the Ricci curvature scalar formed with $ \Gamma^{\lambda}_{\mu\nu} $ the Levi-Civita connection and $ {\cal R}$ is  the Palatini curvature of an independent torsionless connection $ \hat{\Gamma}^{\lambda}_{\mu\nu} $, in analogy with the Palatini approach. Variation of the action \eqref{1} with respect to the metric yields the following gravitational field equations
\begin{equation}\label{2}
	G_{\mu\nu}+F({\cal R}){\cal R} _{\mu\nu}-\frac{1}{2}f({\cal R})g_{\mu\nu}=\kappa T_{\mu\nu},
\end{equation}
where $ F({\cal R}) =\frac{df({\cal R})}{d{\cal R}}$.
Equation \eqref{2} can be rewritten in the following form as

\begin{equation}\label{3}
G_{\mu\nu}=\kappa (T_{\mu\nu}+T^{cur}_{\mu\nu}),
\end{equation}
where
\begin{equation}\label{4}
T^{cur}_{\mu\nu}=\frac{1}{\kappa}\left[\frac{1}{2}f({\cal R})g_{\mu\nu}-F({\cal R} ){\cal R} _{\mu\nu}\right],
\end{equation}
where $ T_{\mu\nu} $ is the stress energy tensor describing the ordinary matter.
We can solve ${\cal R} $ from the trace of the field equations and we get

\begin{equation}\label{5}
2f({\cal R} )-F({\cal R} ){\cal R}=-\kappa T+R.
\end{equation}
\\
To proceed, we consider a homogeneous and isotropic universe with a spatially flat FRW metric as follows
\begin{equation}\label{6}
ds^{2} =-dt^{2}+a^{2}(t)(dx^{2}+dy^{2}+dz^{2}),
\end{equation}
where $ a(t) $  is the scale factor.
The field equations for the gravitational model $ f(R, {\cal R}) $ are obtained as follows \cite{20}

\begin{equation}\label{7}
H^{2}+F{\cal H}^{2}-\frac{1}{6}(F{\cal R}-f)-\frac{\kappa\rho}{3}=0,
\end{equation}

\begin{equation}\label{8}
\frac{\ddot{a}}{a}-{\cal H}^{2}F+\frac{f}{6}+\frac{\kappa}{6}(\rho -3p)=0,
\end{equation}
where
\begin{equation}\label{9}
{\cal H}=\frac{\dot{a}}{a}+\frac{\dot{F}}{2F},
\end{equation}
The acceleration equation \eqref{8} can be written as follows
\begin{equation}\label{10}
\frac{\ddot{a}}{a}={\cal H}^{2}F-\frac{f}{6}-\frac{\kappa}{6}(\rho +3p)=0.
\end{equation}
In the above equation,  $ \ddot{a} $ represents cosmic acceleration, and here the dot denotes a derivative with respect to the cosmic time $ t $.
According to the equations of standard cosmology, the effective energy density $ \rho^{eff}= \frac{3}{\kappa}H^{2} $  and
effective pressure $ p^{eff}= -\frac{1}{\kappa} \left[ 2 \frac{\ddot{a}}{a} + H^{2} \right]$  are defined as follows

\begin{equation}\label{11}
\rho^{eff}=\frac{1}{\kappa}\bigg[\rho+\frac{F({\cal R})-2}{6}R-\frac{f({\cal R})}{2}-\frac{3F^{\prime}({\cal R})^{2}\dot{{\cal R}}^{2}}{2F({\cal R})}+\frac{3F^{\prime\prime}({\cal R})\dot{{\cal R}^{2}}}{2}+\frac{3}{2}HF^{\prime}({\cal R})\dot{{\cal R}}+\frac{3}{2}F^{\prime}({\cal R})\ddot{{\cal R}}\bigg] ,
\end{equation}

\begin{equation}\label{12}
p^{eff}=\frac{1}{\kappa}\bigg[p+\frac{-F({\cal R})+2}{6}R+\frac{f({\cal R})}{2}-\frac{F^{\prime\prime}({\cal R})\dot{{\cal R}^{2}}}{2}-\frac{5}{2}HF^{\prime}({\cal R})\dot{{\cal R}}-\frac{1}{2}F^{\prime}({\cal R})\ddot{{\cal R}}\bigg] ,
\end{equation}

where $ \rho$ and $ p$ are the energy density and pressure of matter, respectively.
By Combining the equations \eqref{11} and \eqref{12}, the following equation is obtained
\begin{equation}\label{13}
\rho^{eff}+p^{eff}=\frac{1}{\kappa}\bigg[\rho+p-\frac{3F^{\prime}({\cal R})^{2}\dot{{\cal R}}^{2}}{2F({\cal R})}+F^{\prime\prime}({\cal R})\dot{{\cal R}}^{2}-HF^{\prime}({\cal R})\dot{{\cal R}}+F^{\prime}({\cal R})\ddot{{\cal R}}\bigg],
\end{equation}
which will be very useful in checking energy conditions in this model.

\section{Energy conditions}

Examining energy conditions to establish limitations on the energy-momentum tensor is highly beneficial in GR theory, particularly in the contexts of gravitational collapse and singularities. The energy conditions can be obtained from the Raychaudhuri equation, which has a geometric structure \cite{21}. Since the GR field equations can relate the Ricci tensor $ R_{\mu\nu} $ to the energy-momentum tensor, by combining the Einstein field equations, we can derive the physical conditions on the energy-momentum tensor \cite{22,23,24}.

\begin{equation}\label{14}
\frac{d\theta}{d\tau}=-\frac{\theta^{2}}{2}-\sigma^{\alpha\beta}\sigma_{\alpha\beta}+\omega^{\alpha\beta}\omega_{\alpha\beta}-R_{\alpha\beta}k^{\alpha}k^{\beta}.
\end{equation}
Here, $\theta  $ denotes the scalar expansion, $\sigma_{\alpha\beta} $ the shear tensor, $ \omega^{\alpha\beta} $ rotation, and $  R_{\mu\nu} $ the Ricci tensor. For $ \sigma^2 = \sigma_{\mu\nu} \sigma^{\mu\nu}\geq 0 $  and $ \omega_{\mu\nu} = 0 $,
$   R_{\mu\nu} k^\mu k^\nu \geq 0  $ the term expresses the condition of gravitational attraction $ (\frac{d\theta}{d\tau}<0) $.
Therefore, the energy conditions, which has a geometric structure, are classified as follows \cite{23}
\begin{equation}\label{15}
 \rm NEC \Longleftrightarrow\rho_{eff} +p_{eff} \geqslant 0,
\end{equation}

\begin{equation}\label{16}
\rm WEC \Longleftrightarrow\rho_{eff} \geq0  \hspace{4mm} and \hspace{4mm} \rho_{eff}+p_{eff} \geqslant 0,
\end{equation}

\begin{equation}\label{17}
\rm SEC\Longleftrightarrow\rho_{eff}+3p_{eff} \geq0  \hspace{4mm} and \hspace{4mm} \rho_{eff}+p_{eff} \geqslant 0,
\end{equation}

\begin{equation}\label{18}
\rm DEC\Longleftrightarrow\rho_{eff} \geq0  \hspace{4mm} and \hspace{4mm} \rho_{eff}\pm p_{eff} \geqslant 0.
\end{equation}
We investigate the energy conditions in the framework of generalized gravity $ f(R, {\cal R}) $ within the FRW cosmology, focusing on the effective field equations. These conditions are formulated in terms of the effective energy density and pressure, which are defined based on the modified Friedmann equations as follows

$\rm NEC $:
\begin{equation}\label{19}
\rho^{eff}+p^{eff}=\frac{1}{\kappa}\bigg[\rho+p-\frac{3F^{\prime}({\cal R})^{2}\dot{{\cal R}}^{2}}{2F({\cal R})}+F^{\prime\prime}({\cal R})\dot{{\cal R}}^{2}-HF^{\prime}({\cal R})\dot{{\cal R}}+F^{\prime}({\cal R})\ddot{{\cal R}}\bigg]\geqslant 0,
\end{equation}

$ \rm WEC $:
\begin{align}\label{20}
&\rho^{eff}=\frac{1}{\kappa}\bigg[\rho+\frac{F({\cal R})-2}{6}R-\frac{f({\cal R})}{2}-\frac{3F^{\prime}({\cal R})^{2}\dot{{\cal R}}^{2}}{2F({\cal R})}+\frac{3F^{\prime\prime}({\cal R})\dot{{\cal R}^{2}}}{2}+\frac{3}{2}HF^{\prime}({\cal R})\dot{{\cal R}}+\frac{3}{2}F^{\prime}({\cal R})\ddot{{\cal R}}\bigg]\geqslant 0 ,&\nonumber\\
&\hspace{4mm}  \hspace{4mm} \rho^{eff}+p^{eff} \geqslant 0,&
\end{align}

$\rm SEC $:
\begin{align}\label{21}
&\rho^{eff}+3p^{eff}=\frac{1}{\kappa}\bigg[\rho+3 p+\frac{-F({\cal R})+2}{3}R+f({\cal R})-\frac{3F^{\prime}({\cal R})^{2}\dot{{\cal R}}^{2}}{2F({\cal R})}+\bigg]\geqslant 0 ,&\nonumber\\
&\hspace{4mm}  \hspace{4mm} \rho^{eff}+p^{eff} \geqslant 0,&
\end{align}

$\rm DEC $:
\begin{align}\label{22}
&\rho^{eff}-p^{eff}=\frac{1}{\kappa}\bigg[\rho-p+\frac{F({\cal R})-2}{3}R-f({\cal R})-\frac{3F^{\prime}({\cal R})^{2}\dot{{\cal R}}^{2}}{2F({\cal R})}+F^{\prime\prime}({\cal R})\dot{{\cal R}^{2}}+4HF^{\prime}({\cal R})\dot{{\cal R}}+2F^{\prime}({\cal R})\ddot{{\cal R}}\bigg]\geqslant 0,&\nonumber\\
&\hspace{4mm}  \hspace{6mm}\rho^{eff}+p^{eff} \geqslant 0, \hspace{6mm} \rho^{eff} \geqslant 0.&
\end{align}

\subsection*{Energy conditions in the gravitational model $ R + \alpha {\cal R}^{n}  $ }

In this subsection, we consider the energy conditions for $ f(R, {\cal R}) $ gravity as proposed below
\begin{equation}\label{23}
f(R,{\cal R})=R+\alpha{\cal R}^{n},
\end{equation}
where $ n $ is an integer constant assumed to be positive. The Ricci scalar can be expressed as a function of the Hubble parameter $ H=\frac{\dot{a}}{a} $ and its derivative as follows
\begin{equation}\label{24}
R=6(2H^{2}+\dot{H}).
\end{equation}
Since inequalities \eqref{19} to \eqref{22} arising from gravitational energy conditions for the $ f(R, {\cal R}) $ model are quite lengthy, for simplicity, we consider only the WEC.
For these types of theories, the constraints of the WEC depend not only on  $n$ and $\alpha$ but also on the observed value of cosmological parameters.
\begin{equation}\label{25}
	G_{\mu\nu}+n\alpha{\cal R} ^{n-1} {\cal R}_{\mu\nu}-\frac{1}{2}\alpha{\cal R}^{n}g_{\mu\nu}=\kappa T_{\mu\nu}.
\end{equation}

To impose energy conditions on $ f(R, {\cal R}) $ gravity, for simplicity, we consider the vacuum case where $ \rho = p = 0$.
The consideration of this assumption does not diminish the generality of the problem, as the energy conditions are satisfied in the ordinary matter component of the energy-momentum tensor (the matter action is minimally coupled to gravity). Consequently, it is crucial to examine the energy conditions pertaining to the geometric component of the energy-momentum tensor. It is important to note that in the investigation of baryonogenesis, we must take into account the ordinary matter component of the energy-momentum tensor. Hence, we will analyze baryonogenesis for the full stress-energy tensor
$T^{eff}_{\mu\nu} = T_{\mu\nu}+T^{cur}_{\mu\nu}$, focusing on regions where $(\partial_{\mu}R)J^{\mu}\neq 0$ (requiring $T_{\mu\nu}\neq 0$ and $\dot{R} \neq 0)$.

Consequently, the WEC for the vacuum scenario, as expressed in equation \eqref{20} regarding the cosmological parameters, can be interpreted as follows

\begin{align}\label{26}
 \rho^{eff}=&\frac{1}{\kappa}\bigg[\frac{\alpha n R}{6}{\cal R}^{n-1}+\frac{(2n-7)}{6(2-n)}R-\frac{3n(n-1)^{2}\dot{{\cal R}^{2}}}{2\alpha(2-n)^{2}}{\cal R}^{-n-1}-\frac{3n(n-1)\dot{{\cal R}^{2}}}{2\alpha(2-n)}{\cal R}^{-n-1}+\frac{3Hn(n-1)\dot{{\cal R}}}{2(2-n)}{\cal R}^{-1}&&\nonumber\\
&+\frac{3n(n-1)\ddot{{\cal R}}}{2(2-n)}{\cal R}^{-1}-\frac{3(n-1)^{2}\dot{{\cal R}}^{2}}{2(2-n)^{2}}{\cal R}^{-1-n}\bigg]\geq 0,
\end{align}

\begin{align}\label{27}
 \rho^{eff}+p^{eff}=&\frac{1}{\kappa}\bigg[-\frac{3n(n-1)^{2}\dot{{\cal R}^{2}}}{2\alpha(2-n)^{2}}{\cal R}^{-n-1}-\frac{n(n-1)\dot{{\cal R}^{2}}}{\alpha(2-n)}{\cal R}^{-n-1}-\frac{Hn(n-1)\dot{{\cal R}}}{(2-n)}{\cal R}^{-1}+\frac{n(n-1)\ddot{{\cal R}}}{(2-n)}{\cal R}^{-1}&&\nonumber\\
&-\frac{(n-1)^{2}\dot{{\cal R}}^{2}}{(2-n)^{2}}{\cal R}^{-1-n}\bigg]\geq 0.
\end{align}
Note that for our study case \eqref{23}, from equation \eqref{5} we can write $ {\cal R} $ in terms of $R$ as follows
\begin{equation}\label{2444}
 {\cal R}=\bigg(\frac{R}{\alpha(n-2)}\bigg)^{\frac{1}{n}}.
\end{equation}
Also we can write
\begin{equation}\label{24444}
 {\cal \dot{R}}=\frac{\dot{R}}{\alpha(n-2)}\bigg(\frac{R}{\alpha(n-2)}\bigg)^{\frac{1}{n}-1},
\end{equation}
\begin{equation}\label{244444}
 {\cal \ddot{R}}=\frac{\ddot{R}}{\alpha n(n-2)}\bigg(\frac{R}{\alpha(n-2)}\bigg)^{\frac{1}{n}-1}+
 \frac{(1-n)\dot{R}^{2}}{\alpha n^{2}(n-2)^{2}}\bigg(\frac{R}{\alpha(n-2)}\bigg)^{\frac{1}{n}-2}.
\end{equation}
By substituting the value of $ {\cal R} $ from equation \eqref{2444}, equations \eqref{26} and \eqref{27} (WEC) can be rewritten as follows for the present universe with values $ H_{0} $, $ R_{0}$ and ${\cal R}_{0}$ as follows, respectively.

\begin{align}\label{28}
& \frac{1}{\kappa}\bigg[\frac{\alpha n R_{0}}{6}\bigg(\frac{R_{0}}{\alpha(2-n)}\bigg)^{\frac{n-1}{n}}+\frac{(2n-7)}{6(2-n)}R_{0}-\frac{3n(n-1)^{2}\dot{{\cal R}_{0}^{2}}}{2\alpha(2-n)^{2}}\bigg(\frac{R_{0}}{\alpha(2-n)}\bigg)^{\frac{-n-1}{n}}-\frac{3n(n-1)\dot{{\cal R}_{0}^{2}}}{2\alpha(2-n)}\bigg(\frac{R_{0}}{\alpha(2-n)}\bigg)^{\frac{-n-1}{n}}&\nonumber\\
&+\frac{3H_{0}n(n-1)\dot{{\cal R}_{0}}}{2(2-n)}\bigg(\frac{R_{0}}{\alpha(2-n)}\bigg)^{\frac{-1}{n}}+\frac{3n(n-1)\ddot{{\cal R}_{0}}}{2(2-n)}\bigg(\frac{R_{0}}{\alpha(2-n)}\bigg)^{\frac{-1}{n}}-\frac{3(n-1)^{2}\dot{{\cal R}_{0}}^{2}}{2(2-n)^{2}}\bigg(\frac{R_{0}}{\alpha(2-n)}\bigg)^{\frac{-n-1}{n}}\bigg]\geq 0,
\end{align}

\begin{align}\label{29}
& \frac{1}{\kappa}\bigg[-\frac{3n(n-1)^{2}\dot{{\cal R}_{0}^{2}}}{2\alpha(2-n)^{2}}\bigg(\frac{R_{0}}{\alpha(2-n)}\bigg)^{\frac{-n-1}{n}}-\frac{n(n-1)\dot{{\cal R}_{0}^{2}}}{\alpha(2-n)}(\frac{R_{0}}{\alpha(2-n)})^{\frac{-n-1}{n}}-\frac{H_{0}n(n-1)\dot{{\cal R}_{0}}}{(2-n)}\bigg(\frac{R_{0}}{\alpha(2-n)}\bigg)^{\frac{-1}{n}}&\nonumber\\
&+\frac{n(n-1)\ddot{{\cal R}_{0}}}{(2-n)}\bigg(\frac{R_{0}}{\alpha(2-n)}\bigg)^{\frac{-1}{n}}-\frac{(n-1)^{2}\dot{{\cal R}_{0}}^{2}}{(2-n)^{2}}\bigg(\frac{R_{0}}{\alpha(2-n)}\bigg)^{\frac{-n-1}{n}}\bigg]\geq 0.
\end{align}
The strong energy condition for this model is also expressed as follows
\begin{align}\label{30}
&\rho_{eff}+3p_{eff}=&\nonumber\\
 &\frac{1}{\kappa}\bigg[\frac{-\alpha n R_{0}}{3}\bigg(\frac{R_{0}}{\alpha(2-n)}\bigg)^{\frac{n-1}{n}}+\frac{(7-2n)}{3(2-n)} R_{0}-\frac{3n(n-1)^{2}\dot{{\cal R}_{0}^{2}}}{2\alpha(2-n)^{2}}\bigg(\frac{R_{0}}{\alpha(2-n)}\bigg)^{\frac{-n-1}{n}}-\frac{6H_{0}n(n-1)\dot{{\cal R}_{0}}}{(2-n)}\bigg(\frac{R_{0}}{\alpha(2-n)}\bigg)^{\frac{-1}{n}}\bigg]\geq 0.
\end{align}
Here, to avoid the length of the above equations, we have not included $\dot{\cal R}_{0}$ and $\ddot{\cal R}_{0}$, which are related to observable parameters through equations \eqref{24444} and \eqref{244444}.

Considering that we have

\begin{equation}\label{31}
	R_0=6H_{0}^{2}(1-q_0),~~~~~\dot{R}=6H_{0}^{3}(j_0-q_0-2),
\end{equation}
\begin{equation}\label{32}
	\ddot{R}_0=6H_{0}^{4}(q_{0}^{2}+8q_0+s_0+6),
\end{equation}
\begin{equation}\label{33}
q=-\frac{1}{H^{2}}\frac{\ddot{a}}{a} ~~~~~~~~~~j=\frac{1}{H^{3}}\frac{\dddot{a}}{a}~~~~~~~~~~~s=\frac{1}{H^{4}}\frac{\ddddot{a}}{a},
\end{equation}
where $q$, $j$ and $s$ denote for the deceleration, jerk and snap parameters, respectively.
To examine the energy conditions in this gravitational model, we plot the WEC and  SEC diagrams numerically as functions of $ n $ for specific values of $ \alpha $ in figures 1 and 2, respectively. Note that in the plots, we utilize the present value of the cosmological parameters $H_0$, $q_{0}$, $j_0$, and $s_{_{0}}$, as reported in Ref.\cite{266}.
\begin{figure*}[ht]
  \centering
  \includegraphics[width=3in]{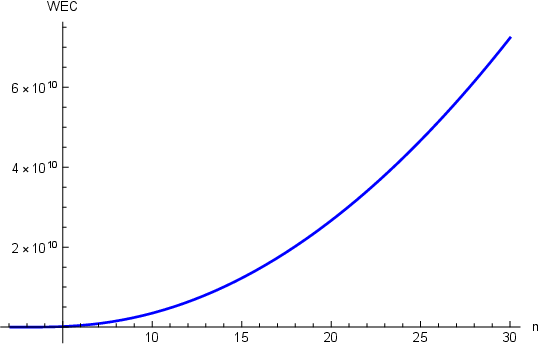}
    \includegraphics[width=3in]{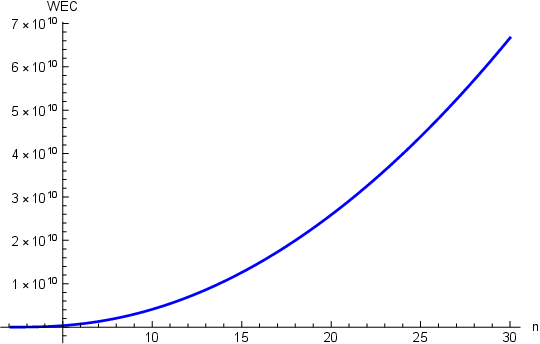}
   \caption{Behavior of the WEC  for $ \rho^{eff}\geqslant 0$ as a function of  $n$, with $  H_{0} = 67.6 ~km s^{-1} Mpc^{-1} $, $ q_{0} = -0.81 $, $j_{0} = 2.16$, $s_{_{0}} = -0.35$, $ \alpha=2 $, and $  \kappa = 1 $, where $ 3 < n < 30 $ (left).
   Behavior of the WEC for $ \rho^{eff}+p^{eff}\geqslant 0 $  as a function of $ n$ (right). }
  \label{stable2}
\end{figure*}

\begin{figure*}[ht]
  \centering
 \includegraphics[width=3in]{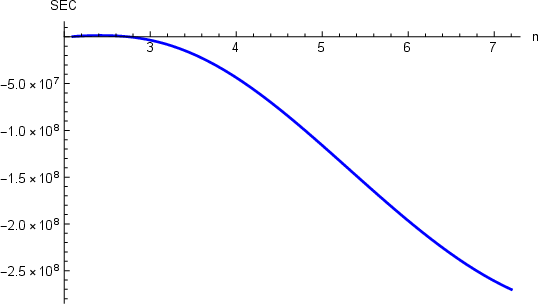}
   \includegraphics[width=3in]{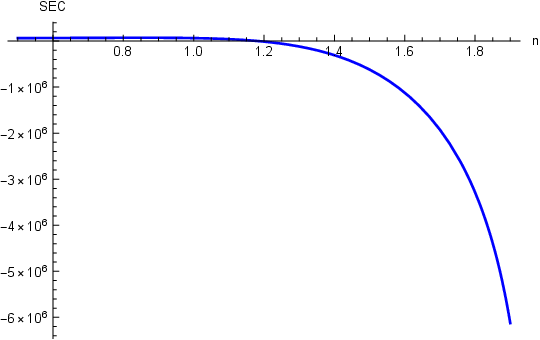}
   \caption{Behavior of the SEC for $ \rho^{eff}+3p^{eff}\geqslant 0 $ as a function of $  n$, with $  H_{0} = 67.6 ~km s^{-1} Mpc^{-1} $, $ q_{0} = -0.81 $, $j_{0} = 2.16$, $s_{_{0}} = -0.35$, $ \alpha=2 $, and $ \kappa = 1 $, where $ 3 < n < 30 $ (left).
   Behavior of the SEC for $ \rho^{eff}+3p^{eff}\geqslant 0 $ as a function of $ n$, where $ 0.5 < n < 1.9 $ (right).}
  \label{stable4}
  \end{figure*}

\section{Field equations of  scalar-tensor theory of $ f(R,{\cal R}) $ }

In this section we start from the following action \cite{25,26}
\begin{equation}\label{36}
	S=\frac{1}{2\kappa}\int d^4{x}\sqrt{-g}f(R,{\cal R})+S_{M}.
\end{equation}
The variation of the action \eqref{36} with respect to the metric and connection, the field equations are obtained in the following form \cite{9}
\begin{flalign}\label{37}
\frac{\partial f}{\partial R}R_{\mu\nu}-\frac{1}{2}g_{\mu\nu}f-(\nabla_{\mu}\nabla_{\nu}-g_{\mu\nu}\square)\frac{\partial f}{\partial R}+\frac{\partial f}{\partial {\cal R}}{\cal R}_{\mu\nu}=\kappa T_{\mu\nu},
\end{flalign}
\begin{equation}\label{38}
\hat{\nabla}_{\lambda}\bigg(\sqrt{-g}\frac{\partial f}{\partial{\cal R}}g^{\mu\nu}\bigg)=0.
\end{equation}
If we consider the action with two scalar fields $\alpha $ and  $\beta  $ as follows \cite{7,255}
\begin{equation}\label{39}
S=\frac{1}{2\kappa}\int d^4{x}\sqrt{-g}\bigg[f(\alpha,\beta)+\frac{\partial f(\alpha,\beta)}{\partial \alpha}
(R-\alpha)+\frac{\partial f(\alpha,\beta)}{\partial \beta}({\cal R}-\alpha)\bigg]+S_{M}.
\end{equation}
It is possible to obtain the field equations by the variation with respect to $ \alpha $ and $ \beta$ from the action of \eqref{36}.
By taking two new auxiliary scalar fields as
\begin{equation}\label{40}
\chi =\frac{\partial f(\alpha,\beta)}{\partial \alpha} \hspace{4mm} and\hspace{4mm} \xi =-\frac{\partial f(\alpha,\beta)}{\partial \beta}.
\end{equation}
The action \eqref{39} can be rewritten as follows
\begin{equation}\label{41}
S=\frac{1}{2\kappa}\int d^4{x}\sqrt{-g}\bigg[(\chi -\xi)R-\frac{3}{2\xi}(\partial\xi)^{2}-V(\chi,\xi)\bigg]+S_{M}.
\end{equation}
Given that the interacting potential $ V(\chi,\xi) $ is defined as follows
\begin{equation}\label{42}
V(\chi,\xi)=-f(\alpha(\chi),\beta(\xi))+\chi\alpha(\chi)-\xi\beta(\xi).
\end{equation}
Now define a new scalar field as $ \phi =\chi-\xi $ and we can perform a conformal transformation to exchange from the Jordan frame to the Einstein frame as follows
\begin{equation}\label{43}
g_{\mu\nu}\rightarrow \tilde{g}_{\mu\nu}=\phi  g_{\mu\nu},
\end{equation}
so, we have
\begin{equation}\label{44}
S=\frac{1}{2\kappa}\int d^4{x}\sqrt{-\tilde{g}}\bigg[\tilde{R}-\frac{3}{2\phi^{2}}(\partial\phi)^{2}-\frac{3}{2\phi\xi}(\partial\xi)^{2}-\frac{W(\phi,\xi)}{\phi^{2}}\bigg]+S_{M}.
\end{equation}
Now, by introducing two new scalar fields as
\begin{equation}\label{45}
\tilde{\phi}=\sqrt{\frac{3}{2}}\frac{\ln\phi}{k}\hspace{4mm}and\hspace{4mm}\tilde{\xi}=\frac{2\sqrt{2}}{\kappa}\sqrt{\xi}.
\end{equation}
Finally, we have
\begin{equation}\label{46}
S=\int d^4{x}\sqrt{-\tilde{g}}\bigg[\frac{1}{2\kappa}\tilde{R}-\frac{1}{2}(\tilde\nabla\tilde{\phi})^{2}-\frac{1}{2}e^{\frac{-\sqrt{2\kappa}\tilde{\phi}}{\sqrt{3}}}
(\tilde\nabla\tilde{\xi})^{2}-\tilde{W}(\tilde{\phi},\tilde{\xi})\bigg]+\tilde{S}_{M},
\end{equation}
\begin{equation}\label{47}
\tilde{S}_{M}=e^{\frac{-\sqrt{2\kappa}\tilde{\phi}}{3}}S_{M},
\end{equation}
where
\begin{equation}\label{48}
\tilde{W}(\tilde{\phi},\tilde{\xi})=\frac{1}{2\kappa}e^{-\frac{\sqrt{2\kappa}\tilde{\phi}}{\sqrt{3}}}W(e^{\frac{\sqrt{2\kappa}
\tilde{\phi}}{\sqrt{3}}},\kappa\frac{\tilde{\xi}^{2}}{8})
\end{equation}
For simplicity, from now on we will omit the tildes in action \eqref{46}.
The field equation obtained by varying action \eqref{46} with respect to $ g_{\mu\nu} $, $ \phi $  and  $ \xi $. The field equations are obtained in the following form
\begin{equation}\label{49}
G_{\mu\nu}=\kappa\bigg(T^{(\phi)}_{\mu\nu}+e^{-\frac{\sqrt{2\kappa}{\phi}}{\sqrt{3}}}(T^{(\xi)}_{\mu\nu}+T_{\mu\nu})- g_{\mu\nu}W
\bigg),
\end{equation}
where $ T^{(\phi)}_{\mu\nu} $ and $ T^{(\xi)}_{\mu\nu} $ define as follows
\begin{equation}\label{50}
T^{(\phi)}_{\mu\nu}=\triangledown_{\mu}\phi \triangledown _{\nu}\phi -\frac{1}{2}g_{\mu\nu}(\nabla\phi)^{2},
\end{equation}
\begin{equation}\label{51}
T^{(\xi)}_{\mu\nu}=\triangledown_{\mu}\xi \triangledown _{\nu}\xi -\frac{1}{2}g_{\mu\nu}(\nabla\xi)^{2}.
\end{equation}
On the other hand, the flat cosmological equations can be written as \cite{27}.
\begin{flalign}\label{52}
3H^{2}=\frac{\kappa}{2}e^{-\frac{\sqrt{2\kappa}{\phi}}{\sqrt{3}}}(\dot{\xi}^{2}-\rho)+\frac{\kappa}{2}\dot{\phi}^{2}+\kappa W,
\end{flalign}
\begin{flalign}\label{53}
2\dot{H}+3H^{2}=-\frac{\kappa}{2}e^{-\frac{\sqrt{2\kappa}{\phi}}{\sqrt{3}}}(\dot{\xi}^{2}-p)-\frac{\kappa}{2}\dot{\phi}^{2}+\kappa W.
\end{flalign}

For a  spatially flat universe, the acceleration equation is given by
\begin{flalign}\label{54}
 \frac{\ddot{a}}{a}=\kappa\bigg[e^{-\frac{\sqrt{2\kappa}{\phi}}{\sqrt{3}}}\frac{(p-\rho)}{2}+\frac{(W-T^{(tot)})}{3}\bigg].
\end{flalign}
We can calculate $T^{(\rm tot)} $ as follows
\begin{equation}\label{55}
T^{(tot)}=T^{(\phi)}+e^{-\frac{\sqrt{2\kappa}{\phi}}{\sqrt{3}}}(T^{(\xi)}+T).
\end{equation}
We know $ T^{(\phi)}=\dot{\phi}^{2} $, $ T^{(\xi)}=\dot{\xi}^{2} $ and $ T=3p-\rho $, so we have
\begin{equation}\label{56}
T^{(tot)}=\dot{\phi}^{2}+e^{-\frac{\sqrt{2\kappa}{\phi}}{\sqrt{3}}}[\dot{\xi}^{2}+3p-\rho].
\end{equation}
Also the the modified Raychaudhuri equation can be written as

\begin{flalign}\label{57}
\frac{\ddot{a}}{a}=\kappa\bigg[\frac{w}{3}-\frac{\dot{\phi}^{2}}{3}-\left(\frac{\dot{\xi}^{2}}{3}+\frac{\rho}{6}+
 \frac{p}{2}\right)e^{-\frac{\sqrt{2\kappa}{\phi}}{\sqrt{3}}}\bigg].
\end{flalign}
According to Friedmann equations in the standard form, the effective energy density and the effective pressure are defined as follows \cite{17}
\begin{equation}\label{58}
\tilde{\rho}^{eff}=\frac{\dot{\phi}^{2}}{2}+e^{-\frac{\sqrt{2\kappa}{\phi}}{\sqrt{3}}}\bigg(\frac{\dot{\xi}^{2}}{2}+\rho -\frac{(W-T^{(tot)})}{3}\bigg),
\end{equation}
\begin{equation}\label{59}
\tilde{p}^{eff}=\frac{\dot{\phi}^{2}}{2}+e^{-\frac{\sqrt{2\kappa}{\phi}}{\sqrt{3}}}\bigg(\frac{\dot{\xi}^{2}}{2}+p +\frac{(W-T^{(tot)})}{3}\bigg).
\end{equation}
By combining equations \eqref{58} and \eqref{59}, the following equation is obtained
\begin{equation}\label{60}
\tilde{\rho}^{eff}+\tilde{p}^{eff}=\dot{\phi}^{2}+e^{-\frac{\sqrt{2\kappa}{\phi}}{\sqrt{3}}}\bigg(\dot{\xi}^{2}+p +\rho\bigg).
\end{equation}
This equation will be very useful in checking energy conditions.

\subsection*{Energy conditions in the scalar-tensor representations}

To consider the energy conditions for the scalar tensor representations of the model in the standard cosmological setting with the effective pressure and  energy density we can write

\begin{equation}\label{61}
\rm NEC:~~~~~~~~~~~~~~~~~~~\rho^{eff}+p^{eff}=\dot{\phi}^{2}+e^{-\frac{\sqrt{2\kappa}{\phi}}{\sqrt{3}}}\bigg(\dot{\xi}^{2}+p +\rho\bigg)\geqslant 0,
\end{equation}

\begin{align}\label{62}
 \rm WEC :~~~~~~~~~&\rho^{eff}=\frac{\dot{\phi}^{2}}{2}+e^{-\frac{\sqrt{2\kappa}{\phi}}{\sqrt{3}}}\bigg(\frac{\dot{\xi}^{2}}{2}+\rho -\frac{(W-T^{(tot)})}{3}\bigg) \geqslant 0,&
\end{align}

\begin{equation}\label{63}
\rm SEC :~~~~~~~~~\rho^{eff}+3p^{eff}=2\dot{\phi}^{2}+e^{-\frac{\sqrt{2\kappa}{\phi}}{\sqrt{3}}}\bigg(2\dot{\xi}^{2}+3p+\rho +\frac{2(W-T^{(tot)})}{3}\bigg) \geqslant 0,
\end{equation}

\begin{align}\label{64}
\rm DEC :~~~~~~~~~~~~~~~~~~~~~\rho^{eff}-p^{eff}=e^{-\frac{\sqrt{2\kappa}{\phi}}{\sqrt{3}}}\bigg(\rho -p -\frac{2(W-T^{(tot)})}{3}\bigg) \geqslant 0.
\end{align}
We take the scalar fields $ \phi $ and $  \xi $ as follows for the simplicity in calculations
\begin{equation}\label{65}
\phi=\alpha a^{n}\hspace{4mm}and\hspace{4mm}\xi=\beta a^{m}.
\end{equation}
By inserting the values $ \dot{\phi} $ and $ \dot{\xi} $ according to equation  \eqref{65} into equations \eqref{56}, \eqref{62} and \eqref{63}, thus
for the current values $ H_{0} $ and $a_{0}=1  $, they can be rewritten as follows
\begin{equation}\label{66}
T^{(tot)}=\alpha^{2}n^{2}H_{0}^{2}a_{0}^{2n}+e^{-\frac{\sqrt{2\kappa}\alpha a_{0}^{n}}{\sqrt{3}}}[\beta^{2}m^{2}H_{0}^{2}a_{0}^{2m}+3p-\rho],
\end{equation}
\begin{align}\label{67}
&\rho^{eff}=\frac{\alpha^{2}n^{2}H_{0}^{2}a_{0}^{2n}}{2}+e^{-\frac{\sqrt{2\kappa}\alpha a_{0}^{n}}{\sqrt{3}}}\bigg(\frac{\beta^{2}m^{2}H_{0}^{2}a_{0}^{2m}}{2}+\rho -\frac{(W-T^{(tot)})}{3}\bigg) \geqslant 0,&
\end{align}
\begin{align}\label{68}
&\rho^{eff}+p^{eff}=\alpha^{2}n^{2}H_{0}^{2}a_{0}^{2n}+e^{-\frac{\sqrt{2\kappa}\alpha a_{0}^{n}}{\sqrt{3}}}\bigg(\beta^{2}m^{2}H_{0}^{2}a_{0}^{2m}+\rho +p\bigg) \geqslant 0,&
\end{align}

\begin{equation}\label{69}
\rho^{eff}+3p^{eff}=2\alpha^{2}n^{2}H_{0}^{2}a_{0}^{2n}+e^{-\frac{\sqrt{2\kappa}\alpha a_{0}^{n}}{\sqrt{3}}}\bigg(2\beta^{2}m^{2}H_{0}^{2}a_{0}^{2m}+3p+\rho +\frac{2(W-T^{(tot)})}{3}\bigg) \geqslant 0.
\end{equation}
Here, we numerically represent the WEC and SEC diagrams  as functions of $  n $ and $  m$ for particular values
of $ \alpha $, $ \beta $ and $W $ to investigate the energy conditions of this gravity model, and present the analysis in the conclusions.

\begin{figure*}[ht]
  \centering
  \includegraphics[width=3in]{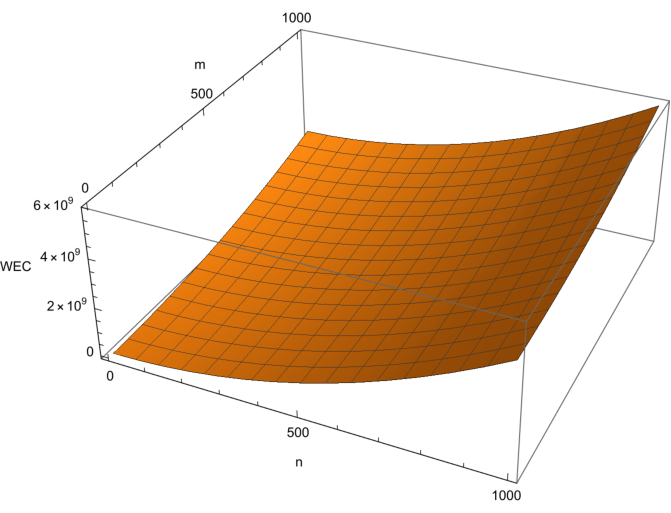}
  \includegraphics[width=3in]{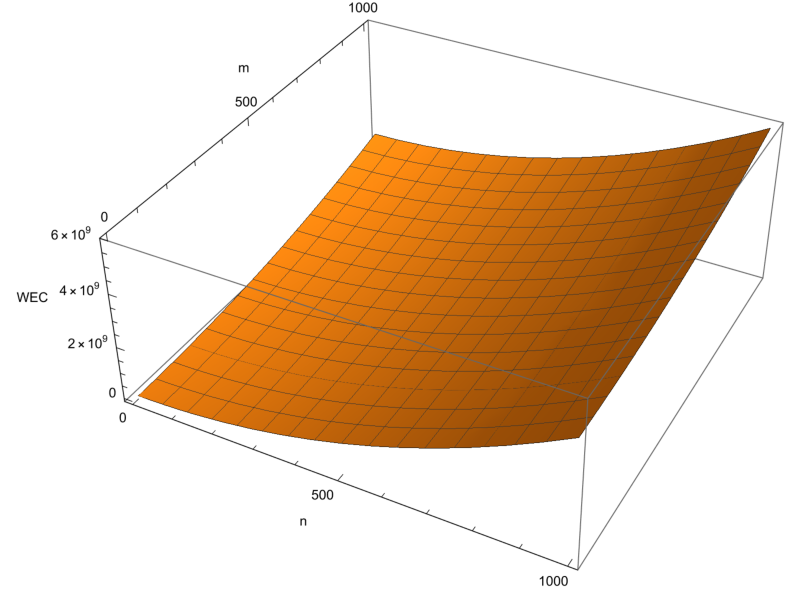}
   \caption{Behavior of the WEC for $ \rho^{eff}\geqslant 0 $ as a function of $ n, m $, with $  H_{0} = 67.6~ km s^{-1} Mpc^{-1} $, $ \rho =p=0 $, $W=1$, $\alpha=\beta=1 $, and $  \kappa = 1 $, where $ 0 < n < 1000 $ and $ 0 < m < 1000 $ (left). Behavior of the  WEC for
   $ \rho^{eff}+p^{eff}\geqslant 0 $ as a function of $ n, m$ for $W=e^{-\frac{\sqrt{2\kappa}\alpha }{\sqrt{3}}}$ (right).}
  \label{stable3}
\end{figure*}


Figure \eqref{stable3} shows that the behavior of weak energy condition for tensor scalar generalized gravity in $ f(R, {\cal R}) $ gravity.


\section{Baryogenesis in $ f(R, {\cal R}) $ gravity }

As mentioned in the introduction, the baryon formation results from the interaction between the Ricci scalar and the baryonic matter flow.
Thus, for the CP-violating interaction of equation (\ref{0}), the corresponding baryon-to-entropy ratio can be written as

\begin{equation}\label{71}
\frac{n_{b}}{s}\simeq \frac{15g_{b}}{4\pi^{2}g_{*}}\frac{\dot{R}}{M^{2}_{*}T_{D}},
\end{equation}
where $n_{b} $ and $ s $ are the number of baryons and the entropy density, respectively, and $ g_{b} $ denotes the entire number of degrees of freedom inalienable to baryons, $ g_{*} $ denotes the whole number of degrees of flexibility of massless particles. In a homogeneous and isotropic universe, when the Ricci scalar is time-varying, this interaction term breaks CPT symmetry.
Interaction term \eqref{0} is a natural and theoretical alternative for the generation of baryon asymmetry in the early universe. This mechanism is based on the intrinsic properties of spacetime within the framework of the GR and does not require any external fields or additional assumptions, which makes it unique and attractive compared to other scenarios \cite{14b}.
Generalized gravitational models such as the teleparallel gravitational model and the $f(R) $ gravitational model are used to explain the baryon production process in the early universe. In these models, the gravitational interaction between gravitons and baryon fields can help explain the anisotropy between baryons and antibaryons without requiring any additional assumptions \cite{12b,14b}.

Another approach to the gravitational baryon formation process is a generalization of the CP-violating interaction, in which instead of coupling the Ricci scalar derivative $(R)$ to the baryon flux, other geometric quantities such as $f(R)$ are used.
These modifications are made within  the framework of modified gravitational theories such as $f(R)$ gravity and geometric models.
This approach is classified under geometric modifications and uses one of the Sakharov criteria that guarantees baryon antibaryon asymmetry due to the presence of CP-violating interactions.
\begin{equation}\label{72}
\frac{1}{M^{2}_{*}}\int \sqrt{-g}\partial_{\mu}f(R)J^{\mu}.
\end{equation}
For a spatially flat FRW background, the $ \frac{n_{b}}{s} $  depends on the time derivative of the Ricci scalar.\\
We aim to investigate the gravitational baryogenesis using other curvature quantities, in particular the hybrid metric-Palatini theory.
In the  $ f(R, {\cal R}) $ gravity, the new gravitational baryogenesis interaction term can be proposed as follows
\begin{equation}\label{73}
\frac{1}{M^{2}_{*}}\int \sqrt{-g}\partial_{\mu}f( R,{\cal R})J^{\mu},
\end{equation}
where $ {\cal R} $ is a Palatini Ricci scalar.
Therefore, in this work we take the gravitational baryogenesis terms corresponding to the $ \partial_{\mu}(f( R, {\cal R})) $ and compare our results with cosmological observations.

The Firedmann equations in the spatially flat universe can be written as

\begin{equation}\label{74}
\dot{H}+H={\cal H}^{2}F-\frac{f}{6}-\frac{1}{6}(\rho +3p),
\end{equation}
where $p=\omega \rho $.

The energy density can be obtained from equations \eqref{11} and \eqref{12} as
\begin{equation}\label{75}
\rho =\frac{6}{1-3\omega}\Big[2H^{2}+\dot{H}+\frac{f}{3}-\frac{F{\cal R}}{6}\Big].
\end{equation}

By taking $f( {\cal R})={\cal R}^{n} $ and performing the mathematical calculations, we get

\begin{equation}\label{76}
\rho =\frac{6}{1-3\omega}\Big[2H^{2}+\dot{H}\Big],
\end{equation}
where from equation (\ref{76}) we can see that the $\rho$ is singular for radiation-dominated universe ($w=1/3$), but in the following we see that in the baryon entropy ratio, {\it i.e.} $n_{b}/s$, baryogenesis can be occurred in the radiation-dominated universe (see equation (\ref{81})).

Also, from equation (\ref{5}) one can write
\begin{equation}\label{77}
 {\cal R}=\Big[\frac{(\rho -3p)+R}{2-n} \Big]^{\frac{1}{n}}.
\end{equation}
Baryon asymmetry requires certain conditions in the early universe for baryons to dominate over anti-baryons.
Based on cosmological observations such as the CMB and BBN, the baryon entropy ratio is given by
\begin{equation}\label{78}
\frac{n_{b}}{s}\simeq 9\times 10^{-11}.
\end{equation}
In the case of CP-violation term in \eqref{73}, the baryon entropy ratio in $f(R, {\cal R})$ gravity is expressed as follows
\begin{equation}\label{79}
\frac{n_{b}}{s}\simeq \frac{15g_{b}}{4\pi^{2}g_{*}}\frac{\dot{R}f_{,R}+\dot{{\cal R}}f_{,{\cal R}}}{M^{2}_{*}T_{D}}.
\end{equation}

To simplify our calculations, let us we assume that $ a(t)=t^{\alpha} $ in which $ \alpha $ is a real constants.
In this case, the value of the Hubble parameter and its derivative is given by
\begin{equation}\label{80}
H=\frac{\alpha}{t},\hspace{4mm} \hspace{4mm}\dot{H}=-\frac{\alpha}{t^{2}},\hspace{4mm} \hspace{4mm}\ddot{H}=\frac{2\alpha}{t^{3}},
\end{equation}
\begin{equation}\label{82}
\dot{R}= \frac{12\alpha(1-2\alpha)}{t^{3}},
\end{equation}

\begin{equation}\label{83}
\dot{\rho}=\frac{12\alpha}{t^{3}(1-3\omega )}\Big[-2\alpha^{2}+\alpha\Big].
\end{equation}

Therefore, the baryon to entropy ratio in equation \eqref{79} for $ f( {\cal R})={\cal R}^{n} $ can be written as follows
\begin{equation}\label{81}
\frac{n_{b}}{s}\simeq \frac{15g_{b}}{4\pi^{2}g_{*}}\frac{[(3-n)\dot{R}+(1-3\omega)\dot{\rho}]}{M^{2}_{*}T_{D}(2-n)}.
\end{equation}
From the above equation it can be seen that for different values of $ \omega $ such as $ \omega= \frac{1}{3} $ (radiation-dominated universe) and $  \omega = 0 $ (matter-dominated universe) baryons can be generated and this value is close to the experimental value given in equation \eqref{78}.
Therefore, the baryon generation problem can be solved in GR if the CP-violating interactions are attributed to metric-Palatini hybridization instead of Ricci scalar. This result shows that using this function instead of the Ricci scalar provides appropriate values for the $\frac{n_{b}}{s}$ that are consistent with cosmological observations.

\begin{figure*}[ht]
  \centering
  \includegraphics[width=2.5in]{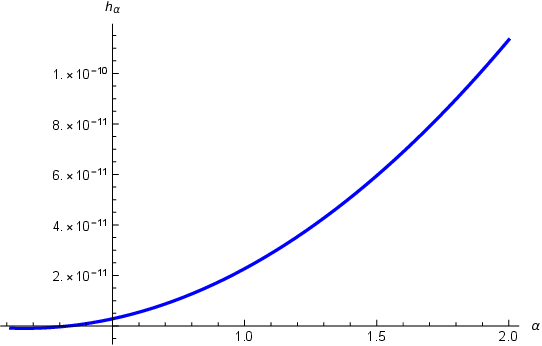}\hspace{1cm}
     \includegraphics[width=2.5in]{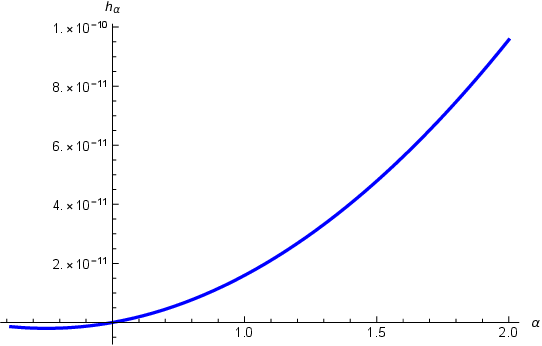}\hspace{1cm}
   \caption{Behavior of the $\frac{n_{b}}{s}  $ as a function of $ \alpha $, with$  M_{*}=9\times 10^{18} GeV$, $ T_D=10^{11}GeV $, $t = 10^{-13} s$, $0.115 <\alpha<2 $, for  $ \omega = 0 $ (left) and $ \omega = \frac{1}{3} $ (right).}
  \label{stable11}
\end{figure*}

Right side figure in \ref{stable11}, for $  \omega = \frac{1}{3} $, which corresponds to the radiation-dominated era of the early universe, is consistent with the values obtained from experimental observations, highlighting the high accuracy of the model in describing the dynamics of the universe and also the left side figure in \ref{stable2} is for $\omega = 0  $, which represents the matter-dominated era, it is observed that baryogenesis still occurs, and the amount of baryon asymmetry during this era is also significant


\section{Conclusions}

$ f(R, {\cal R})$ gravity provides another way to explain the current cosmic acceleration without resorting to the existence of an extra dimension or an exotic component of dark energy. However, the arbitrary choice of different functional forms of $ f(R, {\cal R})$ raises the question of how to limit the many possible theories of $f(R)$ gravity based on physics. In this work, we have shed some light on this issue by discussing some constraints on the general gravity $f(R, {\cal R})$ from the so-called energy conditions. Using the Raychaudhuri equation and the requirement that gravity is an attractive force, we derive zero and strong energy conditions in the $f(R, {\cal R})$ gravitational framework, noting that they are similar to but different from those found in the  general context showed.
\\
In this paper, we studied the energy conditions within the framework of hybrid metric-Palatini gravity. We derived the energy conditions directly from the effective energy-momentum tensor approach under the transformations $\rho\longrightarrow \rho^{eff}$ and $p\longrightarrow p^{eff}$. The inequalities of the energy conditions were calculated in terms of the deceleration and the jerk parameters.
\\
To explore the application of these conditions, we considered a specific form of $f(R, {\cal R})$ gravity, namely  $ f(R, {\cal R}) = R + \alpha {\cal R}^n $. We demonstrated that over a wide range of values of the parameter $n$, the energy conditions can be satisfied. Numerical analysis and the plots confirm that the model is capable of describing an accelerated expansion phase of the universe for many values of $n$, see figures \ref{stable2} for some typical values of $n$.
Additionally, figures \ref{stable4} indicate that the SEC is not satisfied. Therefore, the assumed model predicts the accelerated expansion of the universe, and the validity of the presented model is confirmed through the energy conditions.
\\
In continue we have studied how hybrid metric-Palatini gravity influences gravitational baryogenesis and demonstrates that the baryon asymmetry in the early universe can be satisfactorily explained within this framework. The findings indicate that the baryon-to-entropy ratio obtained in this model is in excellent agreement with experimental data, particularly when considering the effects of CP violation in this gravitational context. The production of gravitational baryons is highly dependent on CP-violating interactions mediated by the Palatini Ricci scalar.
Our results show that the hybrid metric-Palatini gravity model effectively describes the dynamics of the universe across different epochs. Specifically, for both the matter-dominated and radiation-dominated eras, the obtained values are in close agreement with the observed value of approximately
 $10^{-11}$, confirming the models validity as a powerful tool for studying cosmic phenomena.
The consistency between theoretical predictions and observational data suggests that the hybrid metric-Palatini model could play a crucial role in future cosmological studies. Furthermore, the modified baryon-to-entropy ratio derived from this model aligns well with cosmological observations, highlighting its potential to enhance our understanding of baryogenesis and to investigate gravitational processes beyond the framework of standard GR.



\begin{thebibliography}{9}
\bibitem{1}S. Perlmutter, {\it et al.}, astro-ph/9812473.
\bibitem{2}A. G. Riess,  {\it et al.}, Astron. J. 116 (1998) 1009;\\
S. Perlmutter, {\it et al.}, Astrophys. J. 517 (1999) 565.
\bibitem{3} M. Tegmark, {\it et al.}, Phys. Rev. D 69 (2004) 103501.
\bibitem{4}D. N. Spergel,  {\it et al.}, Astrophys. J. Suppl. Ser. 170 (2007) 377.
\bibitem{5}C. Brans  and  R. H. Dicke,  Phys. Rev. 124 (1961) 925;\\
V. Faraoni  ``{\it Cosmology in Scalar Tensor Gravity}'', Dordrecht: Kluwer, (2004);\\
G. Dvali, G. Gabadadze and M. Porrati,  Phys. Lett. B 485 (2000) 20814;\\
T. Jacobson  and D. Mattingly,  Phys. Rev. D 64 (2001) 024028;\\
R. Maartens, Living Rev. Rel. 7 (2004) 7;\\
J. D. Bekenstein,  Phys. Rev. D (2004) 70083509.
\bibitem{6}F. Bombacigno, F. Moretti, G. Montani,  Phys. Rev. D 100 (2019) 124036.
\bibitem{7}J. L. Rosa, S. Carloni, J. P. Lemos,  F. S. Lobo,  Phys. Rev. D, 95(12) (2017) 124035.
\bibitem{8}J. L. Rosa, J. P. Lemos, F. S. Lobo,  Phys. Rev. D 98 (2018) 064054.
\bibitem{9}J. L. Rosa, S. Carloni, J. P.Lemos,   Phys. Rev. D 101 (2020) 104056.
\bibitem{10}J. L. Rosa, J. P. Lemos,   F. S.Lobo,    Phys. Rev. D 101 (2020). 044055.
\bibitem{11}J. L. Rosa,  D. A. Ferreira,  D. Bazeia, F. S. Lobo,  Eur. Phys. J. C 81 (2021) 1.
\bibitem{12}J. L. Rosa, F. S. Lobo, D. Rubiera-Garcia,  JCAP07 (2021) 009.
\bibitem{13}J. L. Rosa,  F. S. Lobo, G. J. Olmo,  Phys. Rev. D 104 (2021) 124030.
\bibitem{14}H. M. R. da Silva, T. Harko, F. S. N. Lobo,  J. L.  Rosa, Phys. Rev. D 104 (2021) 124056.
\bibitem{15}J. L. Rosa,   Phys. Rev. D 104 (2021) 064002.
\bibitem{16}J. L. Rosa,   J. P. Lemos,   Phys. Rev. D 104 (2021) 124076.
\bibitem{17}S. Golsanamlou, K. Atazadeh,  M. Mousavi, Eur. Phys. J. C 83 (2023) 1024.
\bibitem{18}S. M. Carroll, ``{\it An introduction to general relativity: spacetime and geometry}'', Addison Wesley, (2004).
\bibitem{1b} F. Wilczek, Scientific American 243 (1980) 82.
\bibitem{2b}S. Burles, K. M. Nollett, S. M. Turner, Phys. Rev. D 63 (2001) 063512.
\bibitem{3b}D. N. Spergel,  {\it et al.},  Astrophys. J. Suppl. Ser. 148 (2003) 1.
\bibitem{4b}A. G. Cohen, A. De Rujula, S. L. Glashow, Astrophys. J. 495 (1998) 539.
\bibitem{5b}M. Usman,  A. Jawad, and A. M. Sultan, Eur. Phys. J. C 84 (2024)8.
\bibitem{7b}A. D. Sakharov, JETP Lett. 5 (1967) 24.
\bibitem{6b}H. Davoudiasl, R. Kitano, G. D. Kribs, H. Murayama, P. J. Steinhardt, Phys. Rev. Lett. 93, (2004) 201301.
\bibitem{8b}G. Lambiase  and S. Gaetano, Phys. Rev. D 74 (2006) 8.
\bibitem{9b} S. D. Odintsov, and V. K. Oikonomou, Phys. Lett. B 760 (2016) 259.
\bibitem{10b}V. K. Oikonomou, E. N. Saridakis, Phys. Rev. D 94 (2016) 124005;\\
             K. Atazadeh, Eur. Phys. J. C 78 (2018) 455.
\bibitem{11b}M. C. Bento, R. G. Felipe, N. M. C. Santos, Phys. Rev. D 71  (2005) 123517.
\bibitem{12b} M. P. L. P. Ramos, , and J. Paramos, Phys. Rev. D 96 (2017) 10.
\bibitem{13b} S. Bhattacharjee and P. K. Sahoo, Eur. Phys. J. C 80.(2020) 3.
\bibitem{19}S. Carloni, T. Koivisto, and F. S. N. Lobo, Phys. Rev. D 92 (2015) 064035.
\bibitem{20}K. Atazadeh and F. Darabi, Gen. Relat.  Gravi. 46 (2014) 1.
\bibitem{21}A. K. Raychaudhuri,  Phys. Rev. 98 (1955) 1123;\\
 S. Hawking and G. F. R. Ellis, ``{\it The Large Scale Structure of Space-Time''}, Cambridge: Cambridge University Press, (1973);\\
 E. Poisson, {\it ``Advanced general relativity''}, University of Guelph, (2013);\\
 S. Capozziello, T. Harko, T. S. Koivisto, F. S. N. Lobo and G. J. Olmo,  JCAP 04 (2013) 011.
\bibitem{22}D. Rapetti, S. W. Allen, M. A. Amin, and R. D. Blandford, Mon. Not. R. Astron. Soc. 375 (2007) 1510.
\bibitem{23} R. M. Wald, ``{\it General Relativity}'', University of Chicago Press, Chicago (1984).
\bibitem{24}N. J. Popawski,   Class.  Quant. Grav. 24 (2007) 3013.
\bibitem{266}Planck Collaboration, Aghanim, N., Akrami, Y., {\it et al.}, Astro. Astrophy. 641 (2020) A6.
\bibitem{25}N. Tamanini and C. G. Bohmer, Phys. Rev. D 87 (2013) 084031.
\bibitem{26}E. E. Flanagan, Class. Quant. Grav. 21 (2003) 417.
\bibitem{255}C. Gomes, J. L. Rosa and M. A. S. Pinto, arXiv:2506.12870v1.
\bibitem{27}N. Tamanini and C. G. Baohmer, Phys. Rev. D 87 (2013) 084031.
\bibitem{14b}S. Bhattacharjee, Phys. Dark Univ. 30 (2020) 100612.
\end{thebibliography}
\end{document}